\pdfoutput=1

\documentclass[letterpaper, 10 pt, conference]{ieeeconf}  

\IEEEoverridecommandlockouts                              
\usepackage{enumerate}
\usepackage{url}
\usepackage{graphicx}
\usepackage[table]{xcolor}
\usepackage{booktabs}
\usepackage{verbatim}
\makeatletter
\newcommand{\verbatimfont}[1]{\def\verbatim@font{#1}}%
\makeatother
\verbatimfont{\ttfamily\small}
\usepackage{multirow}
\usepackage{array}
\usepackage{amsmath}
\usepackage{centernot}

\usepackage{tabu}
\usepackage{amssymb}
\usepackage{graphics}
\usepackage{mathrsfs}
\usepackage{cite}
\usepackage[mathscr]{euscript}
\usepackage{subcaption}

\newcommand{\aaa}{\mathcal{A}}

\DeclareMathOperator*{\argmax}{arg\,max}

\newcommand{\poa}{\ensuremath{{\rm PoA}}}
\newcommand{\pos}{{\rm PoS}}

\allowdisplaybreaks

\newtheorem{theorem}{Theorem}

\newtheorem{lemma}[theorem]{Lemma}

\title{The Price of Anarchy is Fragile in Single-Selection Coverage Games}
\author{Joshua Seaton and Philip N. Brown
\thanks{This work was supported by the National Science Foundation under grant \#ECCS-2013779.}
\thanks{The authors are with Department of Computer Science at the University of Colorado at Colorado Springs, CO 80918, USA.
	{\tt\small \{jseaton,philip.brown\}@uccs.edu}}}

\begin{document}

\maketitle

\begin{abstract}
    This paper considers coverage games in which a group of agents are tasked with identifying the highest-value subset of resources; in this context, game-theoretic approaches are known to yield Nash equilibria within a factor of $2$ of optimal.
    We consider the case that some of the agents suffer a communication failure and cannot observe the actions of other agents; in this case, recent work has shown that if there are $k>0$ compromised agents, Nash equilibria are only guaranteed to be within a factor of $k+1$ of optimal.
    However, the present paper shows that this worst-case guarantee is \emph{fragile}; in a sense which we make precise, we show that if a problem instance has a very poor worst-case guarantee, then it is necessarily very ``close'' to a problem instance with an optimal Nash equilibrium.
    Conversely, an instance that is far from one with an optimal Nash equilibrium necessarily has relatively good worst-case performance guarantees.
    To contextualize this fragility, we perform simulations using the log-linear learning algorithm and show that average performance on worst-case instances is considerably better even than our improved analytical guarantees.
    This suggests that the fragility of the price of anarchy can be exploited algorithmically to compensate for online communication failures.
\end{abstract}

\section{Introduction}

Game-theoretic approaches to distributed control of multiagent systems have been proposed for such diverse systems as distributed power generation, swarming of autonomous vehicles, network routing, smart grids, and more~\cite{Saad2012,Xu2012,Kordonis2019,Ferguson2020a}.
A common paradigm here involves modeling a distributed optimization problem as a game, endowing each agent with a utility function (an idealized local objective function), and programming each agent to run an algorithm to optimize its own utility function.
This general approach is widely applicable, and leverages a considerable wealth of knowledge from the broader literature on game theory to arrive at generalizable convergence and performance guarantees~\cite{Marden2009a,Gopalakrishnan2010,Tatarenko2014}.

The game-theoretic paradigm holds promise for its generaltiy and modularity~\cite{Marden2014}; however, recent work has suggested that it may lack robustness to unanticipated changes in problem structure~\cite{Brown2017g}.
Accordingly, the past several years have seen a growing interest in the role played by communication and observation among agents in these approaches; for instance, by investigating optimal communication topologies~\cite{Grimsman2018a} and studying the harm induced by communication or observation failures~\cite{Brown2018g,Brown2019c,Jaleel2019a}.

A significant thread of work has studied the application of these game-theoretic approaches for multiagent submodular maximization, a problem with applications in a wide variety of engineering contexts~\cite{Krause2008,Kempe2003,Barinova2012,Lin2011}.
It has long been known that many game-theoretic approaches to this problem are guaranteed to have Nash equilibria always within a factor of $2$ of optimal; it is said that the \emph{price of anarchy} is $1/2$~\cite{Vetta2002}.

Building on this, the recent work in~\cite{Grimsman2020} identifies a class of tight worst-case guarantees for communication-denied multiagent submodular maximization problems.
There, the authors consider a scenario in which some agents cannot observe the actions or detect the contributions of other agents, perhaps due to failed communication or sensing equipment.
In this case, if compromised agents simply ignore the possible presence of other agents (i.e., ``if I can't see you, I'll assume you're not there''), the price of anarchy guarantee worsens from $1/2$ to $1/(k+1)$, where $k$ is the number of compromised agents.
Although it is shown that this bound is tight, the family of worst-case problem instances is finely-tuned; arbitrarily small perturbations to resource values can easily destabilize the low-quality Nash equilibria.
In essence, the specific worst-case examples presented in~\cite{Grimsman2020} are fragile.

In this paper, we show that this fragility is generic, and is in fact a consequence of being worst-case problem instances.
For the class of single-selection coverage games (which are used to generate the worst-case examples in~\cite{Grimsman2020}), we investigate the robustness of the worst-case guarantees offered by~\cite{Grimsman2020} both analytically and empirically.
First, we show the following general principle analytically: if an instance of this game has a very poor price of anarchy, then the game is very \emph{close} (in a sense which we make precise in Section~\ref{sec:model}) to some game with an optimal Nash equilibrium.
On the other hand, any game that is \emph{not} close to some game with an optimal Nash equilibrium necessarily has high-quality Nash equilibria itself.
That is, the price of anarchy for coverage games with compromised agents is fragile.
More precisely, if $G$ is a game with $k>1$ compromised agents that is $D$-\emph{close} to a game with an optimal Nash equilibrium, then the price of anarchy of $G$ satisfies
$$
\poa(G) \geq \frac{1}{1+k-D}.
$$

Subsequently in Section~\ref{sec:sims} we perform an empirical study on the fragility of the price of anarchy by endowing agents with the popular log-linear learning algorithm~\cite{Blume1993,Alos-Ferrer2010,Marden2012a} and comparing the resulting average performance to our analytical worst-case guarantees.
Our simulations suggest that using a payoff-sensitive learning algorithm such as log-linear learning can improve further upon the analytical guarantees, with the greatest relative improvements coming for the game instances with the worst nominal price of anarchy guarantees.
This indicates the possibility that the fragility of the price of anarchy may be exploited algorithmically to offer high performance even in the presence of obstacles to communication.


\section{Model} \label{sec:model}

\subsection{Coverage Game}
A multiagent single-selection set coverage game has agent set $N =\left\{1,\ldots,n\right\}$ and a finite set of resources ${\cal R}$; each resource $r\in{\cal R}$ has value $v_r\in[0,1]$.
We write $v\in\mathbb{R}_{\geq 0}^{|{\cal R}|}$ to represent the set of resource values of a game as a vector.
Each agent $i\in N$ has a set of admissible \emph{actions} given by a subset of resources: $\aaa_i\subseteq {\cal  R}$, and we denote the \emph{joint action space} by $\aaa:=\aaa_1\times\dots\times\aaa_n$.
We require the simple regularity condition that some resource $\bar{r}$ has $v_{\bar{r}}=1$, and that $\bar{r}\in\aaa_i$ for some agent $i$.

We write $a\in\aaa$ to denote a joint action and $a_i$ to denote the resource selected by agent $i$ in $a$.
For a subset of agents $J\subseteq N$, we write $a_J$ to denote the action profile $a$ restricted only to agents in $J$, and accordingly we write $a_{-i}$ to mean  $a_{N \setminus \{i\}}$.
With this notation, we will sometimes write an action profile $a$ as $(a_i,a_{-i})$ or $(a_J, a_{N \setminus J})$ when we wish to highlight the actions of a particular agent or set of agents.
We write $R(a) \subseteq \cal R$ to denote the set of resources selected in the action profile $a$.

In a coverage game, the system-level objective is to maximize the total value of resources covered by agents; that is, to maximize the function
\begin{equation}
    W(a) := \sum_{r\in R(a)} v_r.
\end{equation}

To accomplish this goal, we assume that a system designer has endowed each agent with a \emph{utility function} $U_i:\aaa\to\mathbb{R}$.
Designing these utility functions is a rich area of study~\cite{Paccagnan2019}, but in this manuscript we focus our study by assuming that each agent has nominally been endowed with a \emph{marginal contribution} utility function, given by
\begin{equation} \label{eq:mcdef}
    U_i(a_i, a_{-i}) := W(a_i, a_{-i}) - W(\emptyset, a_{-i}).
\end{equation}
Since each agent is modeled as an individual optimizer that is attempting to maximize its own utility function, the above formulation induces a non-cooperative game.


\subsection{Compromised Agents}

Agents are classified as either \emph{normal} or \emph{compromised}; we write $K\subset N$ to denote the set of compromised agents.
Normal agents receive information about the choices of the other decision-makers within the game.
A compromised agent $i\in K$ receives no information about the choices made by other agents; this models jamming by an adversary, a persistent communication failure, or a failed sensor.
We model this formally via a modified utility function for each agent $i\in K$; we write this modified utility function as
\begin{equation} \label{eq:utilde}
    \Tilde{U}_i(a_i,a_{-i})=v_{a_i}.
\end{equation}
That is, a compromised agent simply assumes that the other (unobserved) agents are not present, and models the value of its choice as the value of the resource it selects.
Note that the modified utility function in~\eqref{eq:utilde} has no dependence on the action of any other agent.
A set coverage game in this context is specified by a tuple $G=(N,{\cal R},A,K,v)$.

With these definitions in hand, the core solution concept we consider is a \emph{pure Nash equilibrium}, or an action profile in which no agent has a unilateral incentive to deviate.
Formally, $a^{ne}$ is a Nash equilibrium if for every normal agent $i\in N\setminus K$ and every $a_i\in\aaa_i$ we have
\begin{equation}
    U_i(a_i^{ne},a_{-i}^{ne}) \geq U_i(a_i,a_{-i}^{ne})
\end{equation}
and for every compromised agent $j\in K$ we have
\begin{equation}
    a_j^{ne} \in \max_{r\in \aaa_j}v_r.
\end{equation}
For game $G$, we write the set of Nash equilibria of $G$ as ${\rm NE}(G)$.%
\footnote{
It can easily be shown that every game of this form has at least one pure Nash equilibrium: once each compromised agent selects a maximum-value resource from its action set, it is well-known that the remaining non-compromised agents simply play a potential game among themselves which is guaranteed to have at least one pure Nash equilibrium.
}

\subsection{Quality of Equilibria}

In a distributed multiagent decision system modeled as a noncooperative game, a system operator is typically tasked with programming each agent with an algorithm that optimizes the agent's utility function; in many formulations, these algorithms are selected to guide the agents collectively to a Nash equilibrium.
Accordingly, the quality of the game's Nash equilibria is a central concern.
In this paper, we measure the worst-case quality of a game's equilibria using the pessimistic measure known as the \emph{price of anarchy}, defined as the worst-case ratio between the objective value of a Nash equilibrium and an optimal action profile, or
\begin{equation}
    \poa(G) = \frac{\min_{a\in {\rm NE}(G)}W(a)}{\max_{a'\in\aaa}W(a')}\leq 1.
\end{equation}
It is known that for any game $G$ as defined in this paper, it holds that if $K\neq \emptyset$, $\poa(G)\geq1/(1+|K|)$~\cite{Grimsman2020}, and that if $K=\emptyset$, $\poa(G)\geq 1/2$~\cite{Vetta2002}.

A companion ``optimistic'' metric is the \emph{price of stability}, defined as the best-case ratio between the objective value of a Nash equilibrium and an optimal action profile, or
\begin{equation}
    \pos(G) = \frac{\max_{a\in {\rm NE}(G)}W(a)}{\max_{a'\in\aaa}W(a')}\leq 1.
\end{equation}
It is well-known that if there are no compromised agents in $G$ (i.e., $K=\emptyset$), then $\pos(G)=1$ for utility functions~\eqref{eq:mcdef} --- that is, there always exists an optimal Nash equilibrium~\cite{Marden2014}.

\subsection{A Notion of Distance from Optimal}

The fundamental goal of this paper is to show that if a game has a very poor price of anarchy, that game is ``close'' in a meaningful sense to a game with an optimal Nash equilibrium (i.e., a game with a price of stability of $1$).
To this end, for game $G=(N,{\cal R},A,K,v)$, let $p\in\mathbb{R}_{\geq0}^{\cal R}$ be a \emph{perturbation vector} of nonnegative numbers to be added to the value vector of $G$.
We write $G_p=(N,{\cal R},A,K,v+p)$ to denote the \emph{perturbed game,} which may have different Nash equilibria and optimal action profiles compared to $G$.
By a correct perturbation to $G$, it may happen that the quality of the Nash equilibria of $G_p$ may be improved relative to $G$.
We define the \emph{distance} of game $G$, denoted $D(G)$, as the minimum total perturbation to $G$ which yields $\pos(G_p)=1$.
Formally, the distance of game $G$ is defined as
\begin{equation} \label{eq:pert}
    D(G) := \inf\left\{|p|_1 : p\in\mathbb{R}_{\geq0}^{\cal R} \mbox{ and }\pos(G_p)=1\right\}
\end{equation}
where $|p|_1$ denotes the $\ell_1$-norm of $p$.
Throughout, we call $p$ an \emph{optimal perturbation vector} if it achieves the infimum in~\eqref{eq:pert}.

\section{Our Contributions}
Previous results~\cite{Grimsman2020} show that a game $G=(N,{\cal R},A,K,v)$ with $|K|$ compromised agents have $\poa(G)\geq \frac{1}{1+|K|}$, in this paper we improve upon this bound that is parameterized by the distance that the game in question is from a game with an optimal Nash equilibrium.
This effectively demonstrates that worst-case instances for this type of game are \emph{fragile} in a strong sense; that is, arbitrarily-small perturbations to their specifications can render their equilibria optimal.

\begin{theorem} \label{thm:main}
Given a game $G$ with distance $D(G)$, it holds that
\begin{equation} \label{eq:main}
    \poa(G)\geq\min\left\{\frac{1}{2},\frac{1}{|K|+1-D(G)}\right\}.
\end{equation}
Furthermore, this bound is tight.
\vspace{2mm}
\end{theorem}

The proof proceeds via a sequence of inequalities; this sequence is adapted to our purposes from one introduced in~\cite{Grimsman2020}.
We repeat it here for completeness, and highlight our modifications to make our novel contribution clear.
First, the proof of Theorem~\ref{thm:main} relies on two new auxiliary lemmas whose proofs appear in the appendix.
These lemmas establish certain useful properties of these games and allow us to regularize our application of the perturbation vector $p$.

\begin{lemma} \label{lem:3}
Without loss of generality, for any optimal action profile $a^{opt}$, it can be assumed that if $r = a_i^{opt}$ for some $i\in K$, then $r\notin A_j$ for any $j\notin K$.
\end{lemma}

Intuitively, Lemma~\ref{lem:3} says that it is unnecessary to consider a certain type of resource conflict between compromised and normal agents. 
In particular, it can be assumed that the actions used by compromised agents in optimal action profiles are not available to non-compromised agents.

\begin{lemma} \label{lem:wlog}
Let ${a}^{opt}$ be an optimal action profile for $G$.
Then there exists an optimal perturbation vector $p$ for $G$ such that if $p_r>0$ for some $r$, then $r=a^{opt}_i$ for some $i\in K$.
\end{lemma}

That is, it can be assumed without loss of generality that an optimal perturbation vector $p$ increases the value \emph{only} of actions that are used in the optimal action profile by compromised agents.
This allows us to significantly reduce the search space for optimal perturbation vectors, and greatly simplifies the forthcoming derivations.

We are now prepared to present the proof of Theorem~\ref{thm:main}.
\subsubsection*{Proof of Theorem~\ref{thm:main}}
Let $G=(N,R,A,K,v)$ be a single-selection resource coverage game with compromised agents with optimal perturbation vector $p$; let $a^{ne}$ denote a  worst-case Nash equilibrium, and $a^{opt}$ denote an optimal action profile for $G$.
In the following, we frequently write $W(a,a')$ to mean $W(a\cup a')$.
First, note that
\begin{align}
    W\left(a^{opt}\right) &\leq W\left(a^{opt},a^{ne}_K\right) \nonumber \\
    &\leq W\left(a^{opt}_{N\setminus K},a^{ne}_K\right) + \sum_{i\in K} W\left(a^{opt}_i\right), \label{eq:pause 1}
    %
    %
\end{align}
where the first inequality is trivial and the second follows from the submodularity of $W$.
Note that Lemmas~\ref{lem:wlog} and~\ref{lem:3} provide that 
\begin{equation}
    \sum_{i\in K} W\left(a^{opt}_i\right) = \sum_{i\in K} W\left(a^{opt}_i\right) - D(G).
\end{equation}

Next, we establish an upper bound on $\sum_{i\in K} W\left(a^{opt}_i\right)$ in two cases.
The first case is when $W(a_K^{ne})<1$.
Here, we have that 
\begin{align}
    \sum_{i\in K} W\left(a^{opt}_i\right) &= \sum_{i\in K} W\left(a^{opt}_i\right) - D(G) \nonumber \\
                        &\leq \sum_{i\in K} W\left(a^{opt}_i\right) - D(G)W(a_K^{ne}) \nonumber \\
                        &\leq W(a_K^{ne})\left(|K| - D(G)\right). \label{eq:first ub}
\end{align}
where the first inequality is $W(a_K^{ne})<1$ and the second is because agents in $K$ can only improve utility by switching to a Nash equilibrium, so that $\sum_{i\in K} W\left(a^{opt}_i\right)\leq \sum_{i\in K} W\left(a^{ne}_i\right)$.

For the second case, let $W(a_K^{ne})\geq1$.
Because $W(a_i)=v_r$ for some $r$, and $v_r\leq1$ for all $r\in{\cal R}$, we have
\begin{align}
    \sum_{i\in K} W\left(a^{opt}_i\right) &= \sum_{i\in K} W\left(a^{opt}_i\right) - D(G) \nonumber \\
                        &\leq |K| - D(G) \nonumber \\
                        &\leq W(a_K^{ne})\left(|K| - D(G)\right). \label{eq:2nd ub}
\end{align}

Combining~\eqref{eq:pause 1},~\eqref{eq:first ub}, and~\eqref{eq:2nd ub}, we have thus far established 
\begin{equation} \label{eq: before cribbing David}
    W\left(a^{opt}\right) \leq W\left(a^{opt}_{N\setminus K},a^{ne}_K\right) + W(a_K^{ne})\left(|K| - D(G)\right).
\end{equation}

From here, following a technique applied in~\cite{Grimsman2020}, it can be shown via a lengthy series of steps that
\begin{equation}
    W\left(a^{opt}_{N\setminus K},a^{ne}_K\right) + W(a_K^{ne}) \leq 2W\left(a_{N\setminus K}^{ne},a_K^{ne}\right). \label{eq:not reprinting David}
\end{equation}
This approach involves explicitly modeling the agents in $N\setminus K$ as playing a ``sub-game'' game among themselves, and leverages the known result that the price of anarchy of a game with no compromised agents is no worse than $1/2$~\cite{Vetta2002}.

Finally, combining~\eqref{eq:not reprinting David} and~\eqref{eq: before cribbing David}, whenever $D(G)\leq|K|-1$, we have
\begin{align}
    W\left(a^{opt}\right)   &\leq 2W\left(a_{N\setminus K}^{ne},a_K^{ne}\right) + W(a_K^{ne})\left(|K| - D(G)-1\right) \nonumber \\
                            &= 2W(a^{ne}) + W(a_K^{ne})\left(|K| - D(G)-1\right) \nonumber \\
                            &\leq W(a^{ne})\left(|K| - D(G)+1\right),\label{eq:theinequality}
\end{align}
where the last inequality follows by the monotonicity of $W(a)$.
Thus,~\eqref{eq:theinequality} illustrates that when $D(G)\leq|K|-1$ it holds that
\begin{equation}
    \frac{W(a^{ne})}{W(a^{opt})} \geq \frac{1}{|K|+1-D(G)},
\end{equation}
illustrating the second case of the minimum in~\eqref{eq:main}.
The first case of the minimum in~\eqref{eq:main} follows from the standard result that when $K=\emptyset$, $\poa(G)\geq1/2$ for the class of valid utility games~\cite{Vetta2002}.

\begin{figure}
    \centering
    \includegraphics[width=.45\textwidth]{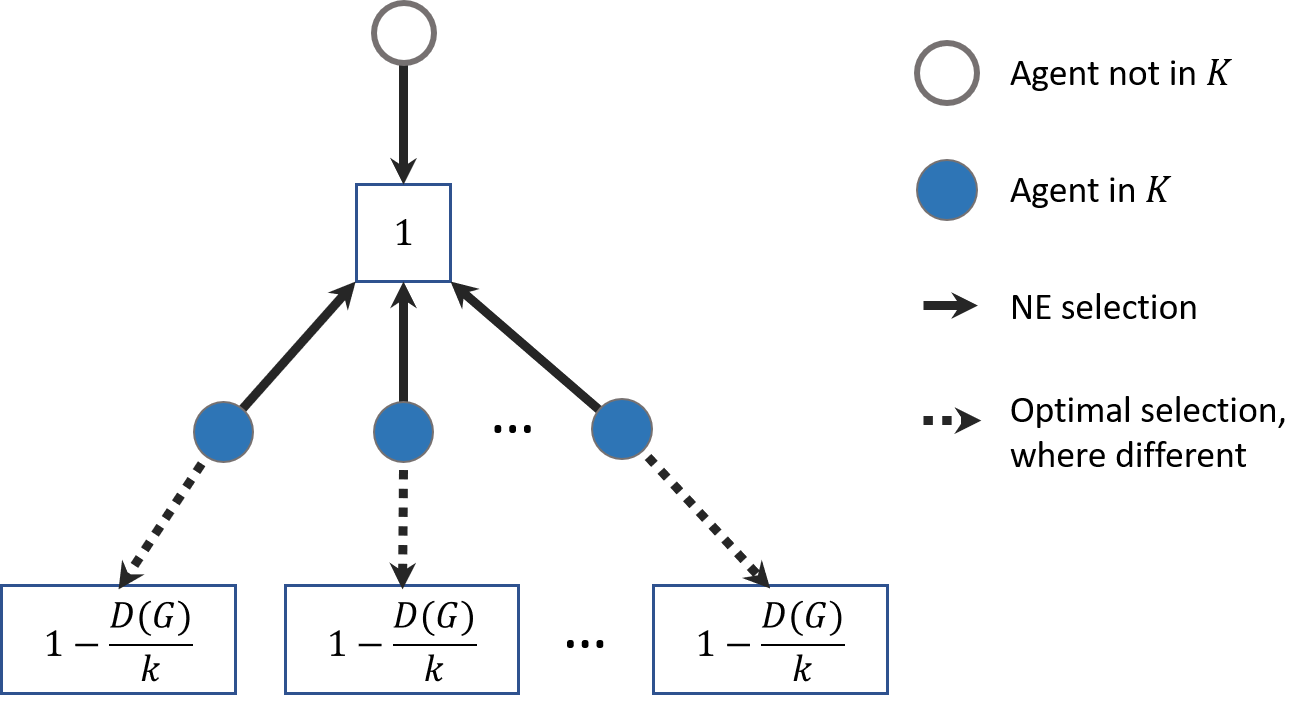}
    \caption{An instance of a single-selection coverage game. It consists of a set of $n=k+1$ agents and $n$ resources. The value of $R_0$ is one. The other members of $\mathcal{R}$ each have a value of $1-\frac{D(G)}{k}$.}
    \label{fig:sim_setup}
\end{figure}

To see that~\eqref{eq:main} is tight, consider the family of examples described in Figure~\ref{fig:sim_setup}; note that whenever $D(G)\leq|K|-1$, the example depicted in~\ref{fig:sim_setup} illustrates tightness.
On the other hand, if $D(G)>|K|-1$, then tightness of the $1/2$ bound follows from standard results, e.g.~\cite{Vetta2002}.
\hfill \QED

\section{Simulations} \label{sec:sims}
In this section we present the results of multiple runs of a simulation of single-selection coverage games. The simulation implements \emph{log-linear learning}. Log-linear learning operates in discrete steps at times $t_0, t_1, ...$, which results in a sequence of joint actions $a(0), a(1), ...$.
It begins with an arbitrary joint action $a(0)$. At each each subsequent step, an agent is chosen randomly and makes a selection based upon its utility function ${U}\left(\right)$.\par
The topology of the simulated game consists of a set of $n$ agents and a set of $r$ resources, the largest of which has a value of one and is designated $R_0$.

\begin{figure}
    \centering
    \includegraphics[width=.52\textwidth]{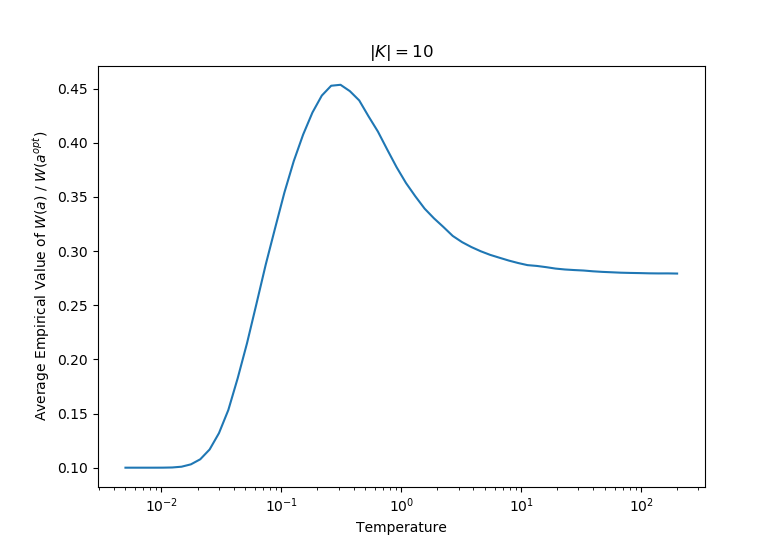}
    \caption{Simulation across a range of temperatures. Although there are $k$ compromised agents, the average objective function value at very low temperatures is $\frac{1}{K}$ because, in this instance, $D(G)=1$, hence ${\rm PoA}(G)=\frac{1}{1 + K - D(G)}=\frac{1}{K}$.}
    \label{fig:across_temps}
\end{figure}

The remaining resources $\mathcal{R}-\{R_0\}$ all have a value $v_{r\in\mathcal{R}-R_0}=1-\frac{D(G)}{k}$ where $k$ is the number of compromised agents and $k=n-1$. All agents have $R_0$ as an admissible action.
The compromised agents also have as an admissible action a unique member of $\mathcal{R}-\{R_0\}$ and a number $z$ of \emph{dummy nodes} as admissible actions. The purpose of the dummy nodes is to decrease the average objective value of simulated game at higher temperatures as their choices become more uniformly random.\par

In Figure~\ref{fig:across_temps} are the results of running the simulation across a range of temperatures. For this simulation, temperatures ranged from $10^{-2.3}$ (approx. $.005$) to $10^{2.3}$ (approx. 199.5). The game that is simulated has the topology described above but with a fixed distance, which, in this case, is arbitrarily $D(G)=1$. There are three basic phases in this plot. At lower temperatures, the agents' selections approximate an asynchronous \emph{best response} to other agents' actions, so the average objective function value for a trial of $200,000$ iterations is $.090887$ of optimal, close to $\frac{1}{1+K-D(G)}=\frac{1}{K}$, which is what we would expect. As the temperature increases, the compromised agents begin to make selections that are less rational but do so in a manner that improves the social welfare, in this specific simulation, to $.45627$ of optimal. As temperatures rise even higher, agents act with increasing irrationality and select coverage of the \emph{dummy nodes}. This behavior is to the detriment of the social welfare. Each agent, both compromised and non-compromised, have three dummy nodes as admissible actions. These nodes are unique to the agent. As the number of dummy nodes is increased, the maximum average social welfare will begin to drop off at lower temperatures and at higher temperatures will move well below the theoretical lower bound of a pure Nash equilibrium.

\begin{figure}
    \centering
    \includegraphics[width=.52\textwidth]{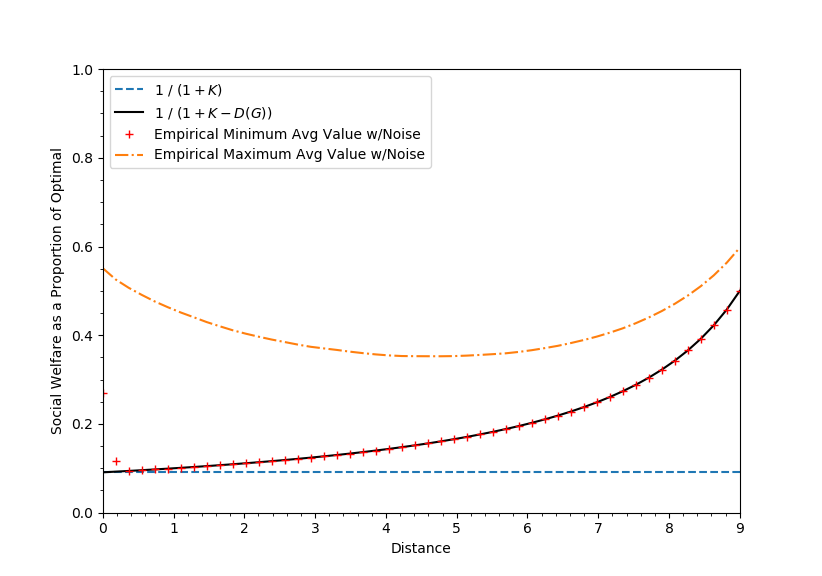}
    \caption{${\rm PoA}\left(G_k\right)=\frac{1}{1+k}$ is the price of anarchy of a coverage game with $k$ compromised agents. }
    \label{fig:verification}
\end{figure}

In Figure~\ref{fig:verification} are the results of running the simulation across a range of temperatures as above but also across distances that range from $0$ to $k-1$. For consistency, we have fixed $|K|=10$. This simulation begins at a fixed \emph{distance}. For a given distance, the simulation is run across a range of temperatures using log-linear learning and performs $200,000$ steps. The minimum and maximum values from this run, similar to the plot in Figure~\ref{fig:across_temps}, are then plotted. The minimum value is represented by the red plus-signs, and the maximum value is represented by the dashed, orange line. It is clear to see that the quality of the social welfare can be greatly increased with a small amount of noise added to the system.\par

For Figure~\ref{fig:fixed_temp}, a large number of simulations were run to find a fixed temperature that maximized the overall quality across a range of \emph{distances}. For this simulation, using a value of $|K|=10$ and with three dummy nodes in the action set of each agent, this temperature was $.55$. Figure~\ref{fig:fixed_temp} illustrates the quality of the average social welfare across distances from $0$ to $|K|-1=9$ with $200,000$ in each iteration.

\section{Conclusion}
While recent work has focused on the lower bounds of efficiency in multiagent systems in which communications failures or other forms of information deprivation can adversely affect performance, this paper demonstrates that these bounds are fragile and that the theoretical lower bound only exists in a small class of single-selection games. Furthermore, probabilistic techniques such as log-linear learning show great promise in guiding system designers to develop techniques to mitigate the potentially poor behavior resulting from communication failure or information deprivation.

\bibliographystyle{ieeetr}
\bibliography{library}

\appendix

\begin{figure}
    \centering
    \includegraphics[width=.52\textwidth]{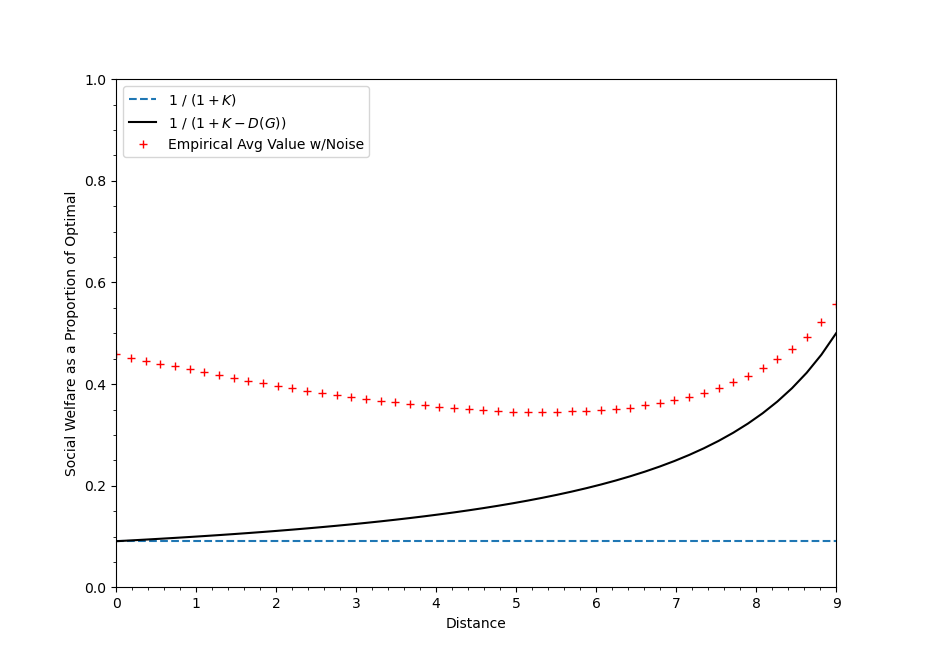}
    \caption{Simulation performed with a fixed \emph{temperature} across a range of distances. This temperature, .55, as a fixed quantity, resulted in the highest average welfare across the range of distances between $0$ and $K-1$.}
    \label{fig:fixed_temp}
\end{figure}

Here we present the proofs of Lemmas~\ref{lem:wlog} and~\ref{lem:3}.

\subsubsection*{Proof of Lemma~\ref{lem:3}}
Given a game $G=(N,{\cal R},A,K,v)$ with a worst-case Nash equilibrium $a^{ne}$, we shall construct a new game $\bar{G}=(N,\bar{{\cal R}},\bar{A},K,\bar{v})$ and show that the worst-case NE of this game $\bar{a}^{ne}$ can be no better than $a^{ne}$, i.e., $W\left(a^{ne}\right) \geq W\left(\bar{a}^{ne}\right)$. Furthermore, $W\left(a^{opt}\right) \leq W\left(\bar{a}^{opt}\right)$; therefore, ${\rm PoA}(G) \geq {\rm PoA}(\bar{G})$.\par

For any resource $r$ such that $i\in K, j\notin K$ and $a^{opt}_{i}=a^{ne}_j=r, j\neq i$, add a new resource $\bar{r}$ such that $\bar{{\cal R}} = {\cal R} \cup \{\bar{r}\}$ and set its value to $\bar{v}_{\bar{r}}=v_r$. If $r$ is an admissible action of any other agent $m\notin K, m\neq i,j$, remove $r$ as an admissible action of $m$ such that $\bar{{\cal A}}_m = {\cal A}_m \setminus \{r\}$. Let $\bar{a}_j=\bar{r}$ and $\bar{{\cal A}}_j = {\cal A}_j \cup \{\bar{r}\} \setminus \{r\}$. Since $a^{ne}$, with the selection of $r$ is a Nash equilibrium of $G$, then, clearly, $\bar{a}^{ne}$, with the selection of $\bar{r}$, is also a Nash equilibrium of $\bar{G}$.\par

The creation of this new game results in an $a^{ne}$ that is guaranteed to be no better than the original. No agent $i\in K$ will be incentivized to unilaterally change its selection from $a^{ne}_i$ of the original game. The removal of the resource $r$ from the action set of only one non-compromised agent and the subsequent addition of a new resource in the same agent's action set cannot improve upon $W\left(a^{ne}\right)$ as the selection of this resource will remain a NE.

Finally, note that $D(G)=D(\bar{G})$; this holds because no compromised agent's action sets or utility functions differ from $G$ to $\bar{G}$, and Lemma~\ref{lem:wlog} shows that an optimal perturbation vector need only modify values of resources accessible by compromised agents.
\hfill\QED

\subsubsection*{Proof of Lemma~\ref{lem:wlog}}
Given a game $G=(N,{\cal R},A,K,v)$, we shall construct a new game $G_p=(N,{\cal R},A,K,v+p)$. 
Let $p$ be an optimal perturbation vector. 
 
The fact that ${\rm PoS}(G_p)=1$ implies that at least one optimal allocation $a^{opt}\in {\cal A}$ is a Nash equilibrium. For any optimal configuration, $a^{opt}$, a compromised agent $j$ will select $a_j^{ne} \in \argmax_{r\in \aaa_j}v_r$. It is necessary to perturb the effective value of the resource $a^{opt}_j=r$ to satisfy $U_j(a^{opt}_j, a^{opt}_{-j}) \geq U_j(a_j, a^{opt}_{-j}), j\in K$.  
Conversely, the strategy of non-compromised agents is already a Nash equilibrium as $U_i(a^{opt}_i, a^{opt}_{-i}) \geq U_i(a_i, a^{opt}_{-i}), i\notin K$. Therefore it is unnecessary to perturb any action not in $a^{opt}$ or an action not in $\aaa_j$ for $j\in K$.
%
\hfill\QED

\end{document}